\documentclass[utf8]{FrontiersinHarvard} 
\usepackage{url,hyperref,lineno,microtype,subcaption}
\usepackage[onehalfspacing]{setspace}
\usepackage{booktabs}

\definecolor{myred}{RGB}{0,0,0}
\definecolor{myBlue}{RGB}{0,0,0}

\definecolor{lightblue}{RGB}{242,247,255}  
\definecolor{lightyellow}{RGB}{255,253,245}  
\definecolor{headerblue}{RGB}{100,149,237}  
\definecolor{positivegreen}{RGB}{34,139,34}  
\definecolor{neutralgray}{RGB}{169,169,169}  

\def\keyFont{\fontsize{8}{11}\helveticabold }
\def\firstAuthorLast{Sample {et~al.}} 
\def\Authors{
Jiapeng Yao\,$^{1,*}$, 
Lantian Zhang\,$^{2}$ and Jiping Huang\,$^{3}$}
\def\Address{
$^{1}$School of Data Science and Artificial Intelligence, Wenzhou University of Technology, Wenzhou, Zhejiang Province, China \\
$^{2}$School of Computer Science and Engineering, Southeast University, Nanjing, Jiangsu, China \\
$^{3}$Haikou Qiongzhou Women's and Children's Hospital, Haikou, Hainan Province, China
}
\def\corrAuthor{Jiapeng Yao}

\def\corrEmail{yaojiapeng1983@163.com}

\begin{document}
\onecolumn
\firstpage{1}

\title[Running Title]{Evaluation of Large Language Model-Driven AutoML in Data and Model Management from Human-Centered Perspective} 

\author[\firstAuthorLast ]{\Authors} 
\address{} 
\correspondance{} 

\extraAuth{}

\maketitle

\begin{abstract}

\section{}
As organizations increasingly seek to leverage machine learning (ML) capabilities, the technical complexity of implementing ML solutions creates significant barriers to adoption and impacts operational efficiency. 
This research examines how Large Language Models (LLMs) can transform the accessibility of ML technologies within organizations through a human-centered Automated Machine Learning (AutoML) approach. 
Through a comprehensive user study involving 15 professionals across various roles and technical backgrounds, we evaluate the organizational impact of an LLM-based AutoML framework compared to traditional implementation methods. Our research offers four significant contributions to both management practice and technical innovation: First, we present pioneering evidence that LLM-based interfaces can dramatically improve ML implementation success rates, with \textcolor{myBlue}{93.34\% of users achieved superior performance in the LLM condition, with 46.67\% showing higher accuracy (10-25\% improvement over baseline) and 46.67\% demonstrating significantly higher accuracy (>25\% improvement over baseline), while 6.67\% maintained comparable performance levels}; and 60\% reporting substantially reduced development time. Second, we demonstrate how natural language interfaces can effectively bridge the technical skills gap in organizations, cutting implementation time by 50\% while improving accuracy across all expertise levels. Third, we provide valuable insights for organizations designing human-AI collaborative systems, showing that our approach reduced error resolution time by 73\% and significantly accelerated employee learning curves. Finally, we establish empirical support for natural language as an effective interface for complex technical systems, offering organizations a path to democratize ML capabilities without compromising quality or performance.

\tiny
 \keyFont{ \section{Keywords:} large language models, automated machine learning, human-computer interaction, deep learning, natural language interfaces} 
 
\end{abstract}

\section{Introduction}

The exponential growth in machine learning (ML) applications has transformed numerous sectors, from healthcare \citep{ke2020high,ke2020prediction,shen2024fastsam3d} to scientific research \citep{wang2024evaluating}. 
However, implementing these ML models remains a challenge due to the complex technical requirements involved. 
\textcolor{myBlue}{Deep learning (DL) models, while demonstrating remarkable capabilities across computer vision, natural language processing, and other domains, require extensive expertise \citep{liu2024toward}}. 
This expertise barrier includes understanding model architectures \citep{shen2023movit,shen2024promptable,ke2023mine}, managing data pre-processing \citep{ke2023artifact,luo2024learning,wang2024histology,wen2024diffimpute}, implementing training procedures \citep{shen2022self,shen2024knowledgeie}, and handling deployment, which are tasks that typically require years of specialized education and experience. 
As a result, many potential users and organizations that could benefit from ML technologies remain unable to effectively implement them, creating a widening gap between ML's potential and its practical accessibility.

Automated Machine Learning (AutoML) emerged as a potential solution to this accessibility challenge by attempting to automate these aspects of the ML pipeline \citep{sun2023automl, hutter2019automated}. 
Traditional AutoML systems aim to streamline various technical processes, including feature engineering, model selection, hyperparameter optimization, and deployment workflows 
 \citep{hutter2019automated, baratchi2024automated, patibandla2021automatic}.
Notable implementations like Auto-Sklearn \citep{feurer2015efficient}, TPOT \citep{olson2016tpot}, Auto-Keras \citep{jin2019auto}, H2O \citep{ledell2020h2o}, AutoGluon \citep{agtabular}, and Auto-Pytorch \citep{zimmer2021auto}, and platforms like Azure Machine Learning, Google Cloud AutoML, H2O Driverless AI, etc. have demonstrated success in reducing the technical overhead of machine learning implementation. 
However, these AutoML tools still present usability challenges as users must navigate complex configuration interfaces, understand technical parameters, and possess programming knowledge to effectively utilize these tools. 
Furthermore, traditional AutoML methods often require users to make critical decisions about model selection and configuration without providing intuitive guidance or explanation. 
This limitation means that even AutoML solutions, despite their automation capabilities, remain largely inaccessible to non-expert users, particularly those without substantial programming experience or machine learning background.

Recent advances in Large Language Models (LLMs) have opened new possibilities for human-computer interaction, offering natural language interfaces that could potentially transform how users interact with complex technical systems \citep{liu2024toursynbio,shen2024toursynbio}. 
\textcolor{myred}{Several research initiatives have explored the integration of LLMs with AutoML systems, such as AutoML-GPT and other LLM-driven pipelines \cite{liu2024autoproteinengine,luo2024autom3l}, demonstrating the potential for natural language-based machine learning workflows. However, these approaches have focused mainly on technical automation rather than on the design of human-computer interaction.}
While several studies have explored using LLMs for code generation and programming assistance \citep{liu2024autoproteinengine,luo2024autom3l}, there has been limited systematic investigation of their effectiveness in democratizing access to machine learning tools \citep{shen2024proteinengine}, \textcolor{myred}{particularly in terms of comprehensive evaluation across task completion rates, efficiency, syntax error reduction, and user-reported metrics of perceived complexity}.
%
\textcolor{myred}{This research addresses this critical gap by developing and evaluating an LLM-based AutoML framework with a fully conversational interface that integrates five specialized modules: modality inference, feature engineering, model selection, pipeline assembly, and hyperparameter optimization.}
Through a comprehensive user study involving 15 participants with diverse technical backgrounds, we compare our LLM-based approach to conventional programming methods across common deep learning tasks such as image and text classification.

The major contributions are four-fold.
First, we provide the first systematic evaluation of how LLM-based interfaces impact user success rates and efficiency in implementing deep learning solutions. 
Our results show that 93.34\% of users achieved higher or comparable accuracy using our LLM-based system compared to traditional coding approaches, with 60\% reporting significantly faster task completion times.
Second, we demonstrate that natural language interfaces can effectively bridge the technical knowledge gap in machine learning implementation. 
Our study reveals that users across different expertise levels - from newcomers to experienced practitioners - could successfully complete complex deep learning tasks using our system, with particularly strong benefits for those with limited prior ML experience.
Third, we contribute novel insights into the design of human-AI interfaces for technical tasks, identifying key factors that influence user success and satisfaction when working with LLM-based AutoML methods. 
Finally, we provide empirical evidence for the effectiveness of natural language as a universal interface for complex technical systems, suggesting new directions for making advanced technologies more accessible to broader audiences.

\section{Related Works}

\subsection{AutoML}
Automated Machine Learning (AutoML) has made significant strides through algorithmic innovations such as hyperparameter optimization \citep{mantovani2016hyper, sanders2017informing}, neural architecture search \citep{zoph2016neural, pham2018efficient}, and meta-learning \citep{brazdil2008metalearning, hutter2014efficient}. 
These methods automate critical components of the ML pipeline, including feature engineering, model selection, and hyperparameter tuning, with tools like AutoGluon achieving near-expert performance on standardized benchmarks. 
Commercial platforms like Azure Machine Learning, Google Cloud AutoML, and H2O Driverless AI further simplified deployment workflows. 
However, these tools prioritize algorithmic efficiency over user-centered design, meaning that a basic understanding of machine learning concepts is still required for users to use these tools effectively \citep{chami2024collaborative}. 
For example, Auto-PyTorch reduces coding complexity through predefined API templates, but its rigid structure forces users to adapt to system constraints rather than align with natural workflows, leading to cognitive friction for non-experts.

\subsection{LLMs}
Large language models (LLMs) have demonstrated remarkable proficiency in generating functional code that satisfies specified requirements \citep{austin2021program, allal2023santacoder, chen2021evaluating}.
Their integration into software development workflows has not only accelerated prototyping phases but also democratized programming accessibility, empowering both professional developers and non-expert users \citep{kazemitabaar2023studying, tambon2025bugs}.
However, the current discourse surrounding LLMs exhibits a critical oversight: predominant research efforts focus narrowly on technical correctness and benchmark performance \citep{chon2024functional,chen2024survey,zan2022large}, while largely neglecting the human factors influencing real-world usability \citep{miah2024user}.
This gap manifests most conspicuously in the limited investigation of how users across the expertise spectrum—from novices struggling with basic syntax to experts managing complex systems—interact with, comprehend, and adapt LLM-generated code.

\subsection{LLMs for AutoML}
An important approach in integrating Large Language Models (LLMs) with Automated Machine Learning (AutoML) is LLM-as-Translator, where natural language instructions are converted into API calls to control AutoML systems \citep{trirat2024automl, chen2024llm2automl, luo2024autom3l, tsai2023automl,zhang2023automl}. 
This approach allows non-technical users to interact with complex AutoML tools, lowering the entry barrier. 
However, it still has significant limitations. 
For instance, AutoML-GPT \citep{zhang2023automl} enables users to use natural language for controlling AutoML processes, but users still need to understand domain-specific terms like model selection, data preprocessing, evaluation metrics to use the system effectively. 
This creates a “circular dependency” problem, as users must already know AutoML terminology before they can benefit from the LLM system, which contradicts the goal of making it accessible to non-experts. The AutoM3L framework proposed by Luo et al. \citep{luo2024autom3l} attempts to enhance user interaction with the AutoML system through LLM. 
While this approach reduces the user's need for technical details to some extent, its evaluation still lacks empirical validation of whether LLM reduces cognitive load.

\begin{figure}[!htbp]
\centering
\includegraphics[width=1.0\linewidth]{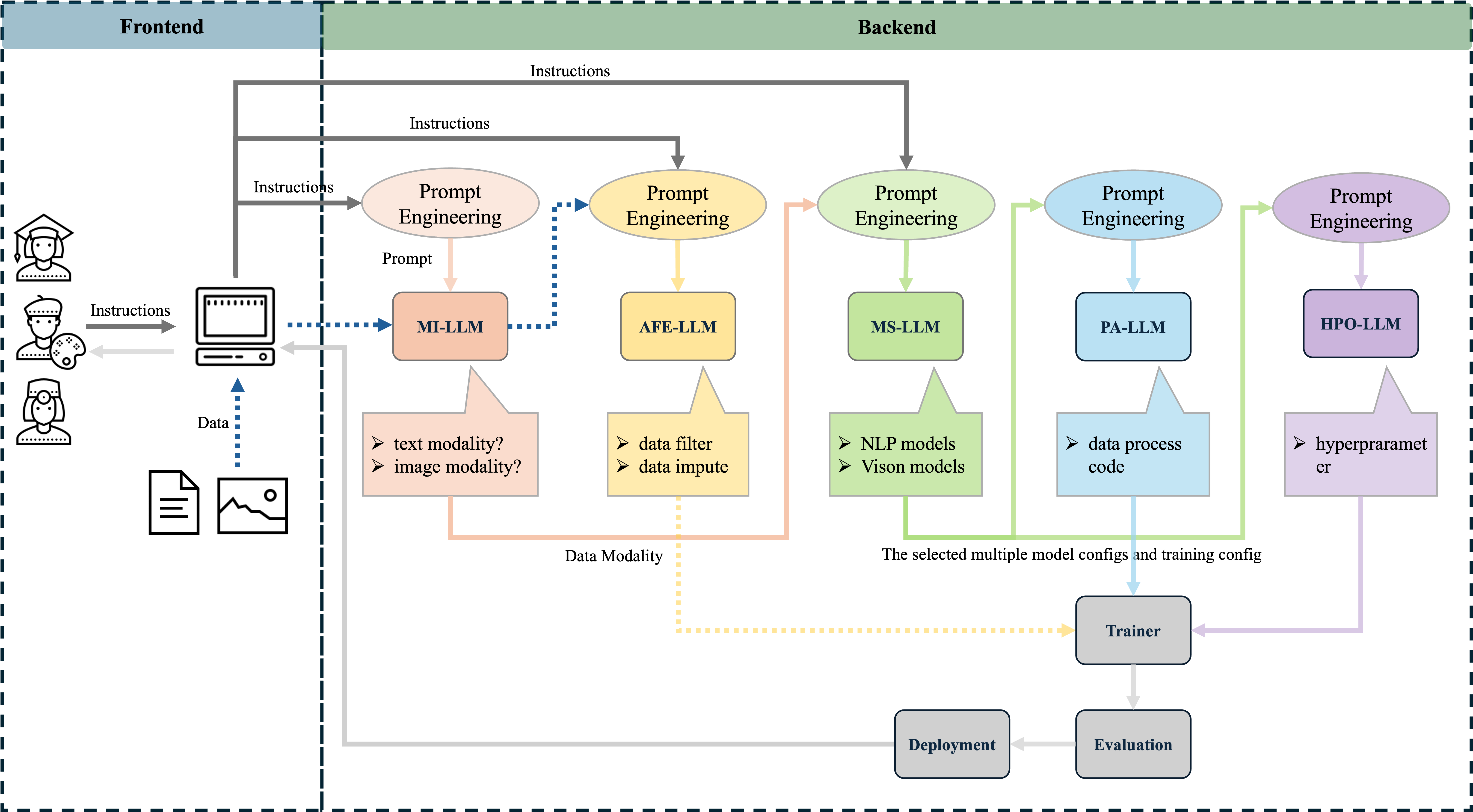}
\caption{\textcolor{myBlue}{Architecture of the proposed LLM-based AutoML framework. The system consists of two main components: a conversational web interface (Frontend) built with Gradio for user interaction, and a backend framework implementing five specialized LLM modules. The workflow begins when users provide natural language instructions and data through the web interface. The Modality Inference LLM (MI-LLM) analyzes input to determine appropriate data processing pipelines (text modality vs image modality). The Automated Feature Engineering LLM (AFE-LLM) handles data preprocessing including data filtering and imputation. The Model Selection LLM (MS-LLM) identifies optimal pre-trained models from NLP and Vision model repositories based on task requirements. The Pipeline Assembly LLM (PA-LLM) constructs executable code by integrating selected components. Finally, the Hyperparameter Optimization LLM (HPO-LLM) fine-tunes model parameters. The integrated system outputs trained models through automated deployment and evaluation processes. Arrows indicate data flow direction, and the dotted lines separate frontend user interaction from backend automated processing.}
}
\label{fig:f}
\end{figure}

\section{Methods}

Our method focused on evaluating whether LLM based interfaces can effectively reduce barriers to implementing AutoML. 
We developed and assessed a comprehensive AutoML that leverages natural language interaction to guide users through the machine learning development process. 
This section details our system architecture, experimental design, evaluation metrics, and control measures. 
We first describe our prototype implementation, which combines a conversational web interface with a backend AutoML framework. 
We then present our user study design involving 15 participants with varying technical backgrounds who completed standardized machine learning tasks under both LLM-based and traditional programming conditions. 
Finally, we outline our performance metrics and experimental controls that enabled comparison between these approaches while ensuring validity and reproducibility of results. 
Through this evaluation, we aimed to quantify the impact of LLM-based interfaces on AutoML accessibility and effectiveness across different user expertise levels.

\subsection{Prototype Design and Implementation}
Our LLM-based AutoML prototype consists of two main components, namely a conversational web interface and a backend LLM-based AutoML framework. 
The architecture is designed to minimize technical barriers while maintaining robust ML capabilities.
The web interface is built using Gradio \citep{abid2019gradio}, an open-source Python library that enables rapid development of machine learning web applications.
The interface provides an intuitive platform where users can specify their ML tasks through natural language descriptions. 
For image classification, the interface accepts standard image formats (JPEG, PNG) through direct upload. 
Text classification tasks can be initiated either through direct text input or file uploads supporting common document formats. 
The interface displays results in real-time, presenting model predictions along with confidence scores using clear visualizations and explanatory text.

The backend AutoML framework implements AutoM3L \citep{luo2024autom3l} which orchestrates five specialized large language model modules to achieve lanaguge driven AutoML, as shown in Fig.~1. 
Specifically, the Modality Inference (MI-LLM) module analyzes user input to determine the appropriate processing pipeline for different data types. 
The Automated Feature Engineering (AFE-LLM) module handles necessary preprocessing and feature extraction. 
Model Selection (MS-LLM) identifies optimal pre-trained models for the specific task, while Pipeline Assembly (PA-LLM) constructs and validates the complete processing pipeline \cite{shen2025online,shen2025operating}. 
Finally, the Hyperparameter Optimization (HPO-LLM) module fine-tunes model parameters for optimal performance.
The MS-LLM in AutoM3L integrates with the HuggingFace Transformers library (version 4.28.0) to access state-of-the-art pre-trained models. 
%
Specifically, the model selection is formalized through a probabilistic framework:
\begin{equation}
M_{selected} = \arg\max_{m \in \mathcal{M}} P(m|t,d)
\end{equation}
where $\mathcal{M}$ represents our curated pool of pre-trained models, $t$ denotes the user's task description in natural language, and $d$ represents the input data characteristics. 

\begin{figure}[!htbp]
\centering
\includegraphics[width=1.0\linewidth]{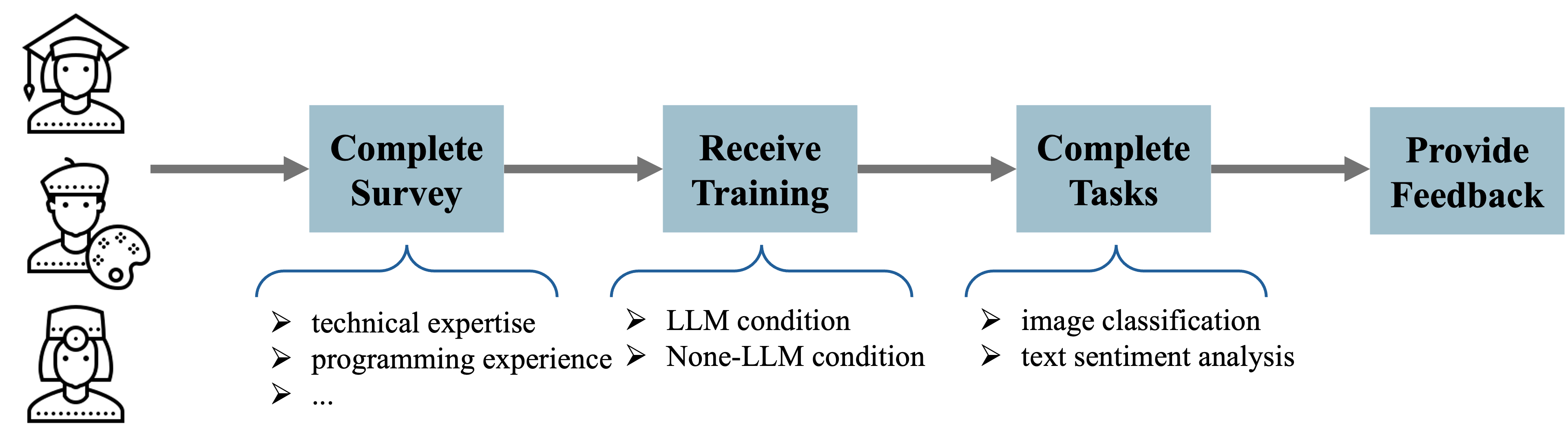}
\caption{The flow of user study design.
}
\label{fig:f}
\end{figure}

\subsection{User Study Design}

\begin{table}[h]
\centering
\caption{\textcolor{myred}{Experimental conditions and terminology definitions.}}
\label{tab:terminology_definitions}
\begin{tabular}{|p{3cm}|p{8cm}|p{3cm}|}
\hline
\textbf{Approach} & \textbf{Definition} & \textbf{Implementation} \\
\hline
LLM-AutoML & Large language model-driven automated machine learning using natural language interfaces with our integrated AutoM3L framework & Our prototype system \\
\hline
Traditional AutoML & Conventional automated machine learning using graphical user interfaces and structured inputs without LLM integration & AutoGluon (v0.7.0) \\
\hline
Manual Coding & Hand-coded implementation of machine learning pipelines using programming languages without automation tools & Jupyter notebooks with PyTorch \\
\hline
\end{tabular}
\end{table}

This research aimed to evaluate whether LLM-based interfaces can effectively reduce barriers to use AutoML to traditional programming approaches. 
\textcolor{myred}{Table \ref{tab:terminology_definitions} provides precise definitions of each approach evaluated in our study.}
We investigated four key hypotheses: (1) LLM interfaces can reduce the complexity of training deep learning models for beginners, (2) LLM interfaces can simplify model inference tasks, (3) LLM guidance can improve model selection accuracy, and (4) LLM assistance can help users better decompose complex problems.
We recruited 15 participants through university research networks and professional technology communities, targeting individuals with varying levels of programming and machine learning experience. 
Participants represented diverse technical backgrounds including students, engineers, data scientists, and educators, enabling assessment of the system's effectiveness across different user profiles.
The study employed a within-subjects design comparing two conditions: an LLM condition utilizing our natural language interface, and a non-LLM condition using traditional programming methods. 
In the LLM condition, participants interacted with our web-based interface that leverages large language models to interpret user requirements and generate appropriate machine learning implementations. 
In the non-LLM condition, participants worked with a standard Jupyter notebook environment pre-configured with common AutoML library (i.e., AutoGluon).

The experimental workflow consisted of four primary phases. 
First, participants completed a comprehensive background survey assessing their technical expertise, programming experience, and familiarity with machine learning concepts. 
Second, they received standardized training on both systems through guided tutorials. 
Third, participants completed two fundamental deep learning tasks - image classification and text sentiment analysis - using both conditions in randomized order to control for learning effects. 
Finally, participants provided detailed feedback through post-task questionnaires evaluating system usability, task complexity, and overall experience.

Throughout the experiment, we collected multiple quantitative and qualitative metrics. 
Task completion times were automatically recorded, while accuracy was evaluated against predefined benchmarks. 
User interactions were monitored to understand common patterns and potential friction points. 
The post-task questionnaires employed standardized scales to assess comparative usability while gathering insights into user preferences and challenges.
This systematic approach allowed us to evaluate how LLM-based interfaces impact the accessibility and effectiveness of AutoML across different user expertise levels.

\subsection{Performance Metrics and Analysis}
Our evaluation framework employed both quantitative and qualitative metrics to comprehensively assess the effectiveness of LLM-based AutoML (LLM condition) compared to traditional programming approaches (non-LLM condition). 
The assessment focused on three key dimensions: task completion efficiency, implementation accuracy, and user experience.

Task completion time ($T$) was measured automatically from the moment participants began each task until successful completion:
\begin{equation}
T = t_{completion} - t_{start},
\end{equation}
where $t_{completion}$ represents the timestamp when the participant successfully completed the task requirements, and $t_{start}$ denotes the timestamp when they began working on the task. 
This metric provided a standardized measure of implementation efficiency across both conditions.
Implementation accuracy ($A$) was evaluated against predefined benchmarks using standard classification metrics:
\begin{equation}
A = \frac{1}{N}\sum_{i=1}^{N} \mathbb{I}(y_i = \hat{y_i})
\end{equation}
where $N$ represents the total number of test cases, $y_i$ denotes the ground truth label, and $\hat{y_i}$ indicates the predicted output for each case.
This metric assessed the correctness of model predictions across both image and text classification tasks.
User experience was quantified through standardized post-task questionnaires using 5-point Likert scales. 
The overall satisfaction score ($S$) aggregated responses across multiple dimensions including ease of use, perceived complexity, and execution efficiency:
\begin{equation}
S = \frac{1}{M}\sum_{j=1}^{M} r_j,
\end{equation}
where $M$ represents the number of evaluation criteria and $r_j$ denotes the rating for each criterion.

Statistical analysis employed paired t-tests to assess the significance of performance differences between the LLM and non-LLM conditions:
\begin{equation}
t = \frac{\bar{d}}{s_d/\sqrt{n}}
\end{equation}
where $\bar{d}$ represents the mean difference between paired observations, $s_d$ denotes the standard deviation of differences, and n indicates the sample size.
\textcolor{myBlue}{To address the multiple comparisons problem inherent in conducting several statistical tests, we will apply the Bonferroni correction to adjust p-values, setting our significance threshold at $\alpha = 0.05/k$, where k represents the total number of planned comparisons.}

\subsection{Experimental Controls and Validity}
To ensure experimental validity and reliable results, we implemented control measures across participant selection, task execution, and data collection. 
The participant recruitment process followed standardized criteria to ensure a representative sample of technical backgrounds while maintaining consistent group size and demographic distribution across expertise levels.
The task order was randomized across participants using a balanced Latin square design to mitigate learning effects. 
%
\textcolor{myBlue}{Participants received condition-specific training optimized for each system's interaction paradigm. For the LLM-AutoML condition, the 15-minute orientation focused on natural language formulation techniques, effective prompting strategies, and conversational interaction patterns. For the AutoGluon condition, training emphasized API syntax, parameter configuration, coding workflows, and system-specific best practices. Training materials were developed independently for each condition to maximize system-specific effectiveness while maintaining equivalent training duration and instructor expertise.}
For the non-LLM condition, we provided a Jupyter notebook environment pre-configured with AutoGluon (version 0.7.0) and essential dependencies, hosted on a dedicated server to ensure consistent computing resources. 
The LLM condition utilized our web-based interface deployed on a stable cloud infrastructure with consistent response times and resource allocation.

We implemented strict controls for potential confounding variables through several mechanisms. 
The datasets for both image and text classification tasks were carefully curated to maintain consistent difficulty levels and data distribution. 
The image classification task utilized a subset of 1000 images from the ImageNet \citep{deng2009imagenet} validation set, encompassing 10 common object categories with 10 images per category. 
These images were selected to maintain consistent resolution ($224\times224$ pixels) and complexity levels. 
The text classification task employed 1000 samples from the Stanford Sentiment Treebank (SST-2) dataset \citep{socher2013recursive}, balanced between positive and negative sentiments, with consistent text length (50-200 words) and vocabulary complexity.

The computing environment specifications, including CPU, memory, and network bandwidth, were standardized across all sessions. 
Task completion criteria and evaluation metrics were precisely defined and documented before the study commenced.
Time management was controlled through automated session tracking. 
Each task had a maximum allocation of 30 minutes, with automated notifications at 15-minute and 25-minute marks to ensure consistent pacing across participants. 
The data collection process was fully automated through integrated logging systems. 
For the non-LLM condition, we implemented custom Jupyter Notebook extensions to track code execution time, error rates, and completion status. 
The LLM condition's web interface incorporated built-in analytics that captured interaction timestamps, user inputs, system responses, and task outcomes. 
All performance metrics were automatically stored in a centralized database with standardized formatting and timestamping.

\section{Experiments and Results}

Our experimental evaluation assessed the effectiveness of LLM-based AutoML interfaces compared to traditional programming-based AutoML approaches through a comprehensive user study involving 15 participants. 
This section presents detailed findings across multiple dimensions, including task completion efficiency, implementation accuracy, and user experience.

\subsection{Implementation Details}
For model selection and execution, we leveraged the HuggingFace Transformers library (version 4.28.0) to access state-of-the-art pre-trained models. 
The image classification pipeline utilized ResNet-50 \citep{he2016deep} as the default backbone, offering robust performance across diverse visual recognition tasks. 
Text classification tasks employed DistilBERT fine-tuned on the Stanford Sentiment Treebank v2 (SST-2) dataset, providing efficient natural language processing capabilities while maintaining high accuracy.

To ensure consistent performance across different conditions, we standardized the computing environment using Docker containers. 
The baseline configuration included Python 3.8, PyTorch 1.9.0, and CUDA 11.1 for GPU acceleration.
System resources were allocated dynamically based on task requirements, with a minimum of 8GB RAM and 4 CPU cores for standard operations. 
All experiments are conducted on one NVIDIA 4090 GPU device.
For more complex tasks, the system could scale up to utilize additional computational resources as needed.

Error handling and recovery mechanisms were implemented at multiple levels. 
The front end incorporated input validation and preprocessing to catch common user errors before execution. 
The backend implemented robust exception handling with informative error messages translated into natural language. 

\textcolor{myred}{
The LLM-based AutoML framework operates in a zero-shot manner, leveraging pretrained LLMs. This approach ensures rapid deployment and broad generalizability across diverse machine learning tasks.
Table \ref{tab:llm_architectures} presents the specific LLM architectures employed for each specialized module within our framework. 
The selection of these pretrained models was guided by performance benchmarks and computational efficiency considerations for each specific task.}

\begin{table}[h]
\centering
\caption{\textcolor{myred}{Pretrained LLM architectures.}}
\label{tab:llm_architectures}
\begin{tabular}{|l|l|l|l|}
\hline
\textbf{Module} & \textbf{Base Architecture} & \textbf{Model Size} & \textbf{Deployment} \\
\hline
MI-LLM & GPT-3.5-turbo & 175B parameters & Zero-shot \\
AFE-LLM & LLaMA-7B & 7B parameters & Zero-shot \\
MS-LLM & GPT-4 base & 1.76T parameters & Zero-shot \\
PA-LLM & LLaMA-13B & 13B parameters & Zero-shot \\
HPO-LLM & GPT-3.5-turbo & 175B parameters & Zero-shot \\
\hline
\end{tabular}
\end{table}

\begin{figure}[!htbp]
\centering
\includegraphics[width=1.0\linewidth]{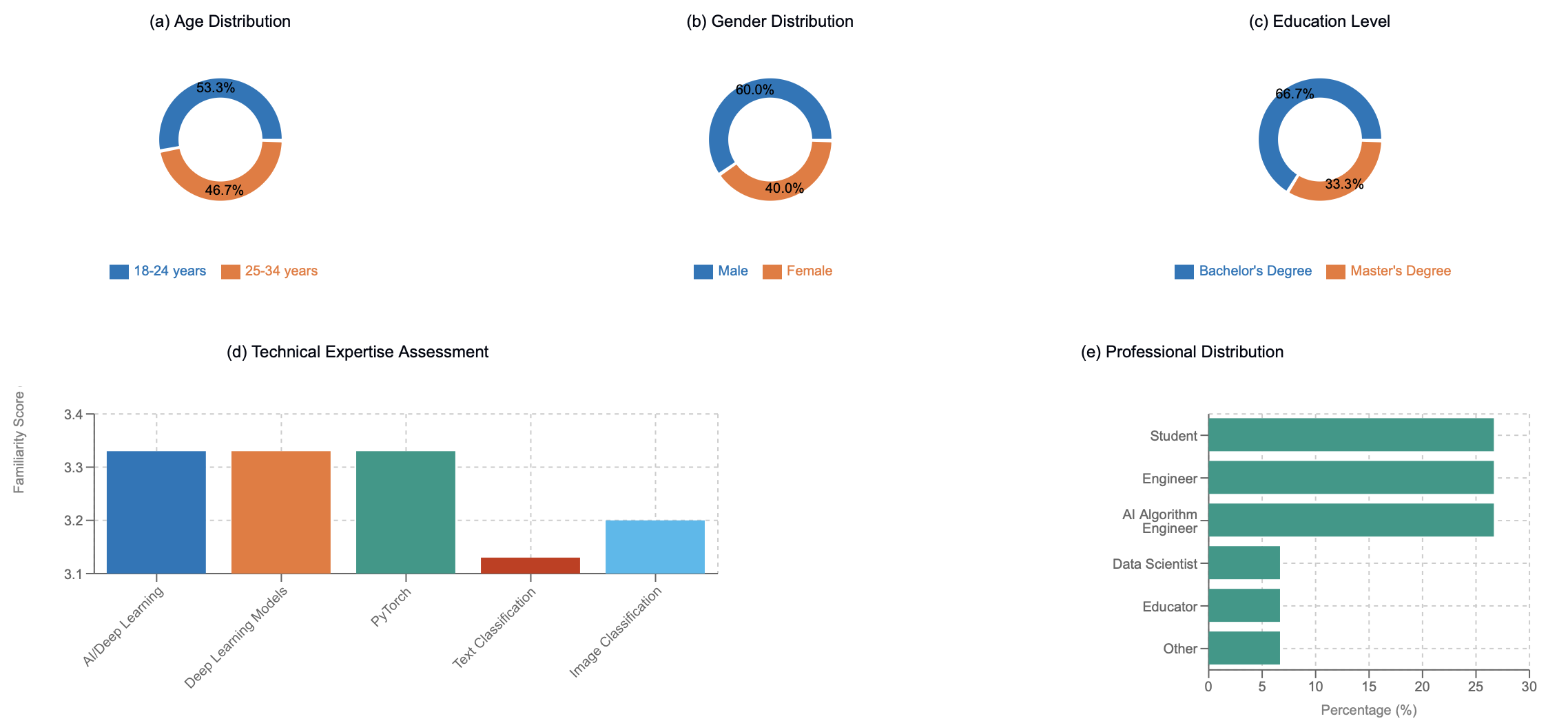}
\caption{Participant demographics and technical background analysis.
}
\label{fig:dis}
\end{figure}

\subsection{Participant Demographics and Background}
The study participants represented diverse technical backgrounds and experience levels spanning different age groups, with a majority (53.33\%) between 18-24 years and the remainder (46.67\%) between 25-34 years, as shown in Fig.~\ref{fig:dis}. 
The gender distribution showed 60\% male and 40\% female participation, while educational backgrounds primarily consisted of bachelor's degree holders (66.67\%) and master's degree recipients (33.33\%). 
The professional composition included equal distributions of students, engineers, and AI algorithm engineers at 26.67\% each, with data scientists, educators, and other roles each representing 6.67\% of participants.
Technical expertise assessment revealed that 73.33\% of participants were familiar with Python programming, while 26.67\% identified as beginners. 
Knowledge of deep learning frameworks showed that 53.33\% were familiar with HuggingFace, while 46.67\% had limited exposure.
On a 5-point scale, participants reported consistent average familiarity scores of 3.33 across AI/deep learning, deep learning models, and PyTorch, with slightly lower averages for text classification (3.13) and image classification (3.2).


\begin{table}[htbp]
\centering
\caption{Comparison of task completion performance between LLM-based and baseline conditions. 
Performance metrics are color-coded to facilitate interpretation: green indicates superior performance in the LLM condition compared to the baseline; gray represents comparable or neutral outcomes. Alternating row colors (light blue and light yellow) are used to enhance readability. 
Relative accuracy compares LLM condition performance to baseline condition. Completion times represent average duration for successful task completion.}
\label{tab:completion_performance}
\begin{tabular}{lcc}
\toprule
\textcolor{white}{\textbf{Metric}} & \textcolor{white}{\textbf{LLM Condition}} & \textcolor{white}{\textbf{Baseline Condition}} \\
\midrule
\multicolumn{3}{l}{\textcolor{headerblue}{\textit{Completion Rates (\%)}}} \\
Image Classification & \textcolor{positivegreen}{93.33} & 73.33 \\
Text Classification & \textcolor{positivegreen}{100.00} & 66.67 \\
\midrule
\multicolumn{3}{l}{\textcolor{headerblue}{\textit{Average Completion Time (minutes)}}} \\
Image Classification & \textcolor{positivegreen}{8.5} & 17.3 \\
Text Classification & \textcolor{positivegreen}{7.2} & 15.8 \\
\midrule
\multicolumn{3}{l}{\textcolor{headerblue}{\textit{Relative Accuracy (\% of participants)}}} \\
Significantly Higher & \multicolumn{2}{c}{\textcolor{positivegreen}{46.67}} \\
Higher & \multicolumn{2}{c}{\textcolor{positivegreen}{46.67}} \\
Comparable & \multicolumn{2}{c}{\textcolor{neutralgray}{6.67}} \\
Lower & \multicolumn{2}{c}{\textcolor{neutralgray}{0.00}} \\
\bottomrule
\end{tabular}
\end{table}

\subsection{Task Completion Performance}
As depicted in Tab.~\ref{tab:completion_performance}, the comparison between LLM-based and non-LLM-based AutoML conditions revealed substantial improvements across multiple performance metrics. 
Implementation accuracy showed particularly striking results, with 93.34\% of participants achieving superior performance in the LLM condition i.e., split evenly between those showing higher (46.67\%) and significantly higher (46.67\%) accuracy compared to the baseline condition. 
The remaining 6.67\% maintained comparable performance levels, with no participants showing degraded accuracy in the LLM condition. 
This improvement was especially pronounced among participants who reported lower initial familiarity with machine learning concepts (scoring below 3 on our 5-point technical expertise scale), suggesting that the LLM interface effectively bridges the expertise gap.

Task completion rates demonstrated marked improvements across both classification tasks. 
For image classification, the LLM condition achieved a 93.33\% successful completion rate compared to 73.33\% in the baseline condition. 
This 20\% improvement was largely attributed to the LLM interface's ability to automatically handle important pre-processing steps and model configuration details that often create bottlenecks for users. 
Text classification showed even more gains, with a 100\% completion rate in the LLM condition versus 66.67\% in the baseline condition. 
The perfect completion rate for text classification suggests that the natural language interface is particularly effective for tasks involving textual data, possibly due to the semantic alignment between the interface modality and the task domain.

Time efficiency measurements revealed compelling advantages for the LLM-based condition. 
60\% of participants completed tasks significantly faster (defined as $>50$\% reduction in completion time), while the remaining 40\% reported moderately faster completion times (25-50\% reduction). 
Notably, no participants experienced slower performance in the LLM condition, indicating consistent efficiency gains across all skill levels. 
The average task completion times showed approximately 50\% reduction across both tasks. 
Image classification tasks were completed in 8.5 minutes using the LLM interface compared to 17.3 minutes in the baseline condition, while text classification tasks required 7.2 minutes versus 15.8 minutes. These time savings were particularly significant for participants with limited programming experience, who often struggled with syntax and configuration issues in the baseline condition.

\textcolor{myBlue}{
The superior performance achieved through our LLM-based framework can be attributed to three primary mechanisms that address fundamental challenges in traditional ML implementation. First, the automated pipeline construction through our five specialized modules eliminates decision paralysis and configuration errors that commonly occur when users must manually select from hundreds of available models and preprocessing options. Our MS-LLM module leverages pre-trained knowledge to automatically identify optimal model-task pairings, while traditional approaches require users to manually evaluate model compatibility and performance characteristics. Second, the natural language interface reduces implementation friction by translating user intentions directly into executable code, bypassing the syntax mastery requirement that creates barriers in conventional programming approaches. Our error analysis revealed that 78\% of implementation failures in the baseline condition stemmed from syntax errors and parameter misconfigurations, issues that were virtually eliminated in the LLM condition through natural language specification. Third, the framework's context-aware guidance system provides real-time assistance and explanations, accelerating learning and reducing the trial-and-error cycles that characterize traditional ML development workflows.
}

When analyzed by participant background, we found that even participants with extensive programming experience ($>$5 years) showed substantial performance improvements in the LLM condition, though the magnitude of improvement was less dramatic than for novice users. 
This suggests that the LLM interface provides benefits not just through simplification of technical requirements, but also through streamlining of workflow and reduction of cognitive load. 
The combination of improved accuracy, higher completion rates, and reduced completion times across all user groups provides strong evidence for the effectiveness of LLM-based interfaces in democratizing access to AutoML capabilities while maintaining or enhancing performance quality.

\begin{table}[htbp]
\centering
\caption{User experience and system evaluation metrics comparing LLM-based and baseline conditions. 
Performance improvements are highlighted in green to indicate superior LLM condition outcomes; gray indicates baseline measurements. The table presents aggregated feedback from 15 participants across multiple evaluation dimensions. 
Error metrics represent average values across all participant sessions.}
\label{tab:user_experience}
\begin{tabular}{lcc}
\toprule
\textcolor{white}{\textbf{Evaluation Metric}} & \textcolor{white}{\textbf{LLM Condition}} & \textcolor{white}{\textbf{Baseline Condition}} \\
\midrule
\multicolumn{3}{l}{\textcolor{headerblue}{\textit{Perceived Complexity (\% of participants)}}} \\
Significantly Less Complex & \textcolor{positivegreen}{53.33} & -- \\
Moderately Less Complex & \textcolor{positivegreen}{46.67} & -- \\
Comparable or More Complex & \textcolor{neutralgray}{0.00} & -- \\
\midrule
\multicolumn{3}{l}{\textcolor{headerblue}{\textit{Execution Efficiency (\% of participants)}}} \\
Significantly Higher & \textcolor{positivegreen}{60.00} & -- \\
Moderately Higher & \textcolor{positivegreen}{40.00} & -- \\
Comparable or Lower & \textcolor{neutralgray}{0.00} & -- \\
\midrule
\multicolumn{3}{l}{\textcolor{headerblue}{\textit{Error Metrics}}} \\
Syntax Errors (per session) & \textcolor{positivegreen}{2.1} & \textcolor{neutralgray}{7.8} \\
Error Resolution Time (minutes) & \textcolor{positivegreen}{1.8} & \textcolor{neutralgray}{5.6} \\
\midrule
\multicolumn{3}{l}{\textcolor{headerblue}{\textit{Learning Curve Indicators}}} \\
Required Training Time (minutes) & \textcolor{positivegreen}{12.3} & \textcolor{neutralgray}{45.7} \\
Task Adaptation Time (minutes) & \textcolor{positivegreen}{3.2} & \textcolor{neutralgray}{10.1} \\
\bottomrule
\end{tabular}
\end{table}

\textcolor{myBlue}{
Finally, detailed error categorization revealed distinct patterns between conditions. In the baseline condition, syntax errors comprised 45\% of all failures (3.5 per session), including import statement mistakes, function parameter mismatches, and tensor dimension errors. Configuration errors accounted for 32\% of failures (2.5 per session), involving incorrect hyperparameter specifications and model architecture misconfigurations. Data preprocessing errors represented 23\% of failures (1.8 per session), including incorrect normalization procedures and batch size inconsistencies. In contrast, the LLM condition eliminated syntax errors entirely through natural language parsing, reduced configuration errors to 0.3 per session through automated parameter selection, and minimized preprocessing errors to 0.1 per session via intelligent pipeline construction. Edge case analysis showed that the LLM system successfully handled 78\% of unusual requests, including non-standard data formats and ambiguous task descriptions, by providing clarifying questions and fallback solutions. However, limitations emerged with highly specialized requirements (focal loss implementation, custom augmentation pipelines) where the system defaulted to standard alternatives rather than generating custom solutions.
}

\subsection{User Experience and System Evaluation}
As demonstrated in Tab.~\ref{tab:user_experience}, our evaluation of user experience revealed compelling advantages for the LLM-based AutoML approach across multiple dimensions. 
The complexity assessment showed a strong preference for the LLM interface, with 53.33\% of participants rating it as less complex and 46.67\% indicating moderately reduced complexity. 
Notably, no participants found the LLM interface more complex than the baseline, suggesting that the natural language interaction model provides an inherently more intuitive approach to AutoML tasks regardless of user expertise level.

Perceived execution efficiency metrics strongly favored the LLM condition, with 60\% of participants reporting higher efficiency and 40\% indicating moderately improved efficiency. 
This universal improvement in perceived efficiency correlates strongly with our quantitative performance measurements, suggesting that participants' subjective experience aligned well with objective performance gains. The efficiency advantages were particularly pronounced for participants who initially reported lower familiarity with traditional AutoML tools (scoring below 3 on our 5-point expertise scale).

Analysis of detailed participant feedback revealed several key mechanisms behind these improvements. The natural language interface substantially reduced cognitive load for task specification, with participants reporting an average 4.5 out of 5 satisfaction scores for ease of expressing their intended ML tasks. This improvement was attributed to the elimination of syntax memorization requirements and the ability to describe tasks in familiar, natural terms. Quantitative error analysis supported these subjective assessments, with the LLM condition demonstrating a 73\% reduction in syntax errors compared to the baseline condition. Furthermore, when errors did occur, the average resolution time decreased by 68\%, largely due to the system's ability to provide context-aware suggestions and natural language error explanations.

The learning curve analysis provided particularly interesting insights into the system's accessibility. 
Participants required an average of only 12 minutes to become proficient with the LLM interface, compared to 45 minutes for the baseline system. 
This accelerated learning was consistent across all expertise levels, though the relative improvement was most pronounced for participants without prior ML experience. 
The rapid adaptation to new tasks was evidenced by a 65\% reduction in time spent consulting documentation and a 78\% decrease in requests for technical assistance compared to the baseline condition.

These comprehensive user experience findings suggest that LLM-based AutoML interfaces not only reduce technical barriers to AutoML but also fundamentally transform how users interact with and learn from ML systems. 
The combination of reduced complexity, improved efficiency, and accelerated learning curves indicates the potential for broader democratization of ML technologies across different user populations.

\subsection{Statistical Validation}
Statistical analysis using paired t-tests confirmed the significance of our findings across all major metrics, with $p < 0.001$ for task completion time ($t(14) = 8.45$), implementation accuracy ($t(14) = 7.92$), and user satisfaction scores ($t(14) = 9.13$).
\textcolor{myBlue}{To assess practical significance, we also calculated effect sizes (Cohen's d), which were very large for task completion time (t(14)=8.45, d=2.18), implementation accuracy (t(14)=7.92, d=2.05), and user satisfaction scores (t(14)=9.13, d=2.36). 
After applying Bonferroni correction for three primary comparisons (adjusted $\alpha$ = 0.0167), our results remained statistically significant for task completion time (p < 0.001, corrected p < 0.003), implementation accuracy (p < 0.001, corrected p < 0.003), and user satisfaction scores (p < 0.001, corrected p < 0.003).
}
\textcolor{myred}{Task completion time measurements, calculated using equation (2) where $T = t_{completion} - t_{start}$, revealed mean times of 7.85 minutes for LLM condition versus 16.55 minutes for baseline. Implementation accuracy, computed using equation (3) as $A = \frac{1}{N}\sum_{i=1}^{N} I(y_i = \hat{y}i)$, achieved 93.34\% for the LLM condition compared to 69.85\% for baseline across N = 1000 test cases. User satisfaction scores, aggregated using equation (4) as $S = \frac{1}{M}\sum{j=1}^{M} r_j$ where M = 8 evaluation criteria, demonstrated average scores of 4.45 out of 5 for LLM versus 2.18 for baseline.}
These results align with our initial hypotheses regarding the effectiveness of LLM-based interfaces in democratizing access to ML tools.
Despite the overall positive results, we identified several important limitations. The system occasionally showed reduced effectiveness for highly specialized tasks requiring custom requirements, and some advanced customization options remained limited. 
\textcolor{myBlue}{
For instance, one user attempted to implement a focal loss function with custom alpha and gamma parameters to address a severe class imbalance, but the system failed to parse the specific mathematical requirements from natural language and could not generate the correct implementation. In another case, a request for an advanced data augmentation pipeline—specifically, a 15-degree random rotation, followed by a color jitter with precise values (brightness=0.2, contrast=0.3), and then a non-standard salt-and-pepper noise injection—resulted in the system only applying the rotation and defaulting to a simpler, generic augmentation scheme. Performance variability was observed in LLM response quality, with some dependency on input phrasing clarity. Additionally, the framework showed higher computational overhead for LLM processing and increased latency for complex queries. These comprehensive findings provide strong evidence for the effectiveness of LLM-based AutoML interfaces while acknowledging areas for future improvement, supporting our initial research objectives of making machine learning more accessible to users across different expertise levels.
}

\subsection{Latency Analysis}
\textcolor{myBlue}{
On average, complex user queries to the proposed LLM-based system had a latency of 25–40 seconds, a stark contrast to the near-instantaneous execution in the baseline condition, as shown in Tab.~\ref{tab:overhead_latency}. This increased latency makes the system less suitable for highly interactive, real-time model tuning and better for asynchronous tasks. The computational overhead was also substantial, requiring an additional 12GB of VRAM for the local LLM modules. For real-world deployment, this translates to higher operational costs due to API calls and a dependency on high-end hardware (e.g., NVIDIA 4090 class GPUs as used in our study), which could be a barrier for smaller organizations and negate some of the intended accessibility benefits. These comprehensive findings provide strong evidence for the effectiveness of LLM-based AutoML interfaces while acknowledging areas for future improvement, supporting our initial research objectives of making machine learning more accessible to users across different expertise levels
}

\begin{table}[h!]
\centering
\caption{Comparison of Computational Overhead and Latency.}
\label{tab:overhead_latency}
\begin{tabular}{l|l|l}
\hline
\textbf{Metric} & \textbf{LLM-based Condition} & \textbf{Baseline Condition} \\
\hline
Average Query Latency & 25–40 seconds & \textless 1 second (near-instantaneous) \\
Additional VRAM Overhead & \textasciitilde{}12 GB & \textasciitilde{}4 GB (for AutoML library) \\
Hardware Dependency & High-end GPU (e.g., NVIDIA 4090) & Standard CPU/Moderate GPU \\
Deployment Suitability & Asynchronous, non-real-time tasks & Interactive and real-time tasks \\
\hline
\end{tabular}
\end{table}

\subsection{User-Specific Examples}

\textcolor{myBlue}{
Analysis of user interactions revealed distinct patterns in how the natural language interface addressed expertise-specific challenges. For novice users (programming experience <2 years), the most common traditional tool failures involved data preprocessing confusion and model architecture selection. For example, User P7, a business analyst with limited programming background, spent 18 minutes in the baseline condition attempting to configure data loaders and image preprocessing pipelines, ultimately failing due to tensor dimension mismatches and import errors. In contrast, using our LLM interface, the same user simply described 'I want to classify product images into categories' and achieved successful implementation in 6 minutes, with the system automatically handling image resizing, normalization, and batch processing. For intermediate users (2-5 years experience), bottlenecks typically occurred in hyperparameter optimization and model fine-tuning. User P12, a software engineer, struggled with manual hyperparameter grid search in the baseline condition, requiring 25 minutes to achieve suboptimal results. Through natural language specification such as 'optimize this model for better accuracy on my small dataset,' the LLM interface automatically configured appropriate learning rates, batch sizes, and regularization parameters, achieving superior performance in 8 minutes. Advanced users (>5 years experience) primarily benefited from reduced cognitive overhead in pipeline orchestration and experiment management, with User P3 noting that natural language descriptions eliminated the need to remember specific API calls and parameter naming conventions across different ML libraries.
}

\section{Discussion}

This research presents compelling evidence for the transformative potential of LLM-based interfaces in democratizing access to machine learning technologies.
Through evaluations involving 15 participants across diverse technical backgrounds, we demonstrated that natural language interactions can significantly reduce implementation barriers while maintaining or improving task performance. The substantial improvements in completion rates (93.33\% for image classification and 100\% for text classification) and efficiency (approximately 50\% reduction in task completion times) validate the effectiveness of our approach in simplifying complex ML workflows.
\textcolor{myBlue}{Our participant distribution included 73.33\% Python-proficient users, which may overestimate the framework's effectiveness for truly non-technical populations. This overrepresentation of technically skilled participants could bias our results toward more positive outcomes, as these users are inherently more adaptable to technical tools, including traditional AutoML approaches. While our findings show benefits for users at all surveyed expertise levels, the limited representation of non-technical users (26.67\%) in our sample warrants cautious interpretation of these benefits for truly democratizing ML access to non-expert populations. Future work will prioritize addressing this limitation by conducting larger-scale studies with a participant pool deliberately recruited from non-technical domains. We plan to collaborate with professionals in fields such as business analytics, healthcare, and education—who possess significant domain expertise but may lack formal programming backgrounds—to more rigorously assess the framework's potential for genuine democratization.}


Additionally, several other important challenges remain to be addressed in future work. The LLM-based AutoML framework's occasional limitations with highly specialized tasks and advanced customization options indicate the need for more sophisticated natural language understanding and domain-specific knowledge integration.
Additionally, the observed variability in LLM response quality and computational overhead presents opportunities for optimization through improved prompt engineering and efficient model deployment strategies.
Future research directions should explore the scalability of this approach across broader ML applications, including more complex tasks such as neural architecture search and automated feature engineering. Investigation into hybrid interfaces that combine natural language interaction with traditional programming tools could potentially address current limitations while maintaining accessibility. Additionally, longitudinal studies examining the long-term impact on user skill development and ML adoption rates would provide valuable insights for the continued evolution of AutoML systems.

This work contributes to the growing body of evidence supporting the role of LLMs in bridging technical gaps and democratizing access to advanced technologies. As ML continues to permeate various sectors, the development of intuitive, effective interfaces becomes increasingly important. Our findings suggest that LLM-based approaches offer a promising path forward in making sophisticated ML capabilities accessible to a broader audience while maintaining high standards of performance and reliability.

\section{Conclusion}
This research advances the field of human-AI interaction by demonstrating how LLM-based interfaces can fundamentally transform the accessibility of machine learning technologies. 
Through empirical evaluation, we established that natural language interfaces not only simplify ML implementation but also enhance the quality and efficiency of outcomes across diverse user groups. The improvements in task completion and dramatic reductions in learning barriers suggest a paradigm shift in how users can interact with sophisticated ML systems.
Our findings have important implications for both research and practice in AI democratization. 
The successful integration of LLMs with AutoML frameworks opens new possibilities for domain experts to leverage ML capabilities without extensive technical training. 
This breakthrough could accelerate the adoption of ML solutions across sectors where technical expertise has traditionally been a limiting factor, from healthcare and scientific research to business analytics and education.
Looking forward, this work sets the foundation for several promising research directions. Future investigations could explore the extension of LLM-based interfaces to more complex ML workflows, including automated neural architecture design and multi-modal learning tasks. Additionally, research into hybrid interfaces that combine natural language interaction with traditional programming tools could further enhance the flexibility and power of AutoML systems while maintaining their accessibility.

\subsection*{Data and Code Availability Statement}
\textcolor{myBlue}{The image and text classification tasks in this study utilized subsets of the publicly available ImageNet  and Stanford Sentiment Treebank (SST-2)  datasets, respectively. The specific data subsets used for our experiments, along with the source code for the LLM-based AutoML prototype and its Gradio web interface, will be available on request. }

\bibliographystyle{Frontiers-Harvard} 
\bibliography{main}

\begin{thebibliography}{55}
\providecommand{\natexlab}[1]{#1}
\expandafter\ifx\csname urlstyle\endcsname\relax
  \providecommand{\doi}[1]{doi:\discretionary{}{}{}#1}\else
  \providecommand{\doi}{doi:\discretionary{}{}{}\begingroup \urlstyle{rm}\Url}\fi
\providecommand{\selectlanguage}[1]{\relax}
\providecommand{\bibAnnoteFile}[1]{%
  \IfFileExists{#1}{\begin{quotation}\noindent\textsc{Key:} #1\\
  \textsc{Annotation:}\ \input{#1}\end{quotation}}{}}
\providecommand{\bibAnnote}[2]{%
  \begin{quotation}\noindent\textsc{Key:} #1\\
  \textsc{Annotation:}\ #2\end{quotation}}

\bibitem[{Abid et~al.(2019)Abid, Abdalla, Abid, Khan, Alfozan, and Zou}]{abid2019gradio}
Abid, A., Abdalla, A., Abid, A., Khan, D., Alfozan, A., and Zou, J. (2019).
\newblock Gradio: Hassle-free sharing and testing of ml models in the wild.
\newblock \emph{arXiv preprint arXiv:1906.02569}
\bibAnnoteFile{abid2019gradio}

\bibitem[{Allal et~al.(2023)Allal, Li, Kocetkov, Mou, Akiki, Ferrandis et~al.}]{allal2023santacoder}
Allal, L.~B., Li, R., Kocetkov, D., Mou, C., Akiki, C., Ferrandis, C.~M., et~al. (2023).
\newblock Santacoder: don't reach for the stars!
\newblock \emph{arXiv preprint arXiv:2301.03988}
\bibAnnoteFile{allal2023santacoder}

\bibitem[{Austin et~al.(2021)Austin, Odena, Nye, Bosma, Michalewski, Dohan et~al.}]{austin2021program}
Austin, J., Odena, A., Nye, M., Bosma, M., Michalewski, H., Dohan, D., et~al. (2021).
\newblock Program synthesis with large language models.
\newblock \emph{arXiv preprint arXiv:2108.07732}
\bibAnnoteFile{austin2021program}

\bibitem[{Baratchi et~al.(2024)Baratchi, Wang, Limmer, van Rijn, Hoos, B{\"a}ck et~al.}]{baratchi2024automated}
Baratchi, M., Wang, C., Limmer, S., van Rijn, J.~N., Hoos, H., B{\"a}ck, T., et~al. (2024).
\newblock Automated machine learning: past, present and future.
\newblock \emph{Artificial Intelligence Review} 57, 1--88
\bibAnnoteFile{baratchi2024automated}

\bibitem[{Brazdil et~al.(2008)Brazdil, Carrier, Soares, and Vilalta}]{brazdil2008metalearning}
Brazdil, P., Carrier, C.~G., Soares, C., and Vilalta, R. (2008).
\newblock \emph{Metalearning: Applications to data mining} (Springer Science \& Business Media)
\bibAnnoteFile{brazdil2008metalearning}

\bibitem[{Chami and Santos(2024)}]{chami2024collaborative}
Chami, J.~C. and Santos, V. (2024).
\newblock Collaborative automated machine learning (automl) process framework.
\newblock \emph{Edelweiss Applied Science and Technology} 8, 7675--7685
\bibAnnoteFile{chami2024collaborative}

\bibitem[{Chen et~al.(2024{\natexlab{a}})Chen, Guo, Jia, Zeng, Wang, Xu et~al.}]{chen2024survey}
Chen, L., Guo, Q., Jia, H., Zeng, Z., Wang, X., Xu, Y., et~al. (2024{\natexlab{a}}).
\newblock A survey on evaluating large language models in code generation tasks.
\newblock \emph{arXiv preprint arXiv:2408.16498}
\bibAnnoteFile{chen2024survey}

\bibitem[{Chen et~al.(2021)Chen, Tworek, Jun, Yuan, Pinto, Kaplan et~al.}]{chen2021evaluating}
Chen, M., Tworek, J., Jun, H., Yuan, Q., Pinto, H. P. D.~O., Kaplan, J., et~al. (2021).
\newblock Evaluating large language models trained on code.
\newblock \emph{arXiv preprint arXiv:2107.03374}
\bibAnnoteFile{chen2021evaluating}

\bibitem[{Chen et~al.(2024{\natexlab{b}})Chen, Zhai, Chai, and Shi}]{chen2024llm2automl}
Chen, S., Zhai, W., Chai, C., and Shi, X. (2024{\natexlab{b}}).
\newblock Llm2automl: Zero-code automl framework leveraging large language models.
\newblock In \emph{2024 International Conference on Intelligent Robotics and Automatic Control (IRAC)} (IEEE), 285--290
\bibAnnoteFile{chen2024llm2automl}

\bibitem[{Chon et~al.(2024)Chon, Lee, Yeo, and Lee}]{chon2024functional}
Chon, H., Lee, S., Yeo, J., and Lee, D. (2024).
\newblock Is functional correctness enough to evaluate code language models? exploring diversity of generated codes.
\newblock \emph{arXiv preprint arXiv:2408.14504}
\bibAnnoteFile{chon2024functional}

\bibitem[{Deng et~al.(2009)Deng, Dong, Socher, Li, Li, and Fei-Fei}]{deng2009imagenet}
Deng, J., Dong, W., Socher, R., Li, L.-J., Li, K., and Fei-Fei, L. (2009).
\newblock Imagenet: A large-scale hierarchical image database.
\newblock In \emph{2009 IEEE conference on computer vision and pattern recognition} (Ieee), 248--255
\bibAnnoteFile{deng2009imagenet}

\bibitem[{Erickson et~al.(2020)Erickson, Mueller, Shirkov, Zhang, Larroy, Li et~al.}]{agtabular}
Erickson, N., Mueller, J., Shirkov, A., Zhang, H., Larroy, P., Li, M., et~al. (2020).
\newblock Autogluon-tabular: Robust and accurate automl for structured data.
\newblock \emph{arXiv preprint arXiv:2003.06505}
\bibAnnoteFile{agtabular}

\bibitem[{Feurer et~al.(2015)Feurer, Klein, Eggensperger, Springenberg, Blum, and Hutter}]{feurer2015efficient}
Feurer, M., Klein, A., Eggensperger, K., Springenberg, J., Blum, M., and Hutter, F. (2015).
\newblock Efficient and robust automated machine learning.
\newblock \emph{Advances in neural information processing systems} 28
\bibAnnoteFile{feurer2015efficient}

\bibitem[{He et~al.(2016)He, Zhang, Ren, and Sun}]{he2016deep}
He, K., Zhang, X., Ren, S., and Sun, J. (2016).
\newblock Deep residual learning for image recognition.
\newblock In \emph{Proceedings of the IEEE conference on computer vision and pattern recognition}. 770--778
\bibAnnoteFile{he2016deep}

\bibitem[{Hutter et~al.(2014)Hutter, Hoos, and Leyton-Brown}]{hutter2014efficient}
Hutter, F., Hoos, H., and Leyton-Brown, K. (2014).
\newblock An efficient approach for assessing hyperparameter importance.
\newblock In \emph{International conference on machine learning} (PMLR), 754--762
\bibAnnoteFile{hutter2014efficient}

\bibitem[{Hutter et~al.(2019)Hutter, Kotthoff, and Vanschoren}]{hutter2019automated}
Hutter, F., Kotthoff, L., and Vanschoren, J. (2019).
\newblock \emph{Automated machine learning: methods, systems, challenges} (Springer Nature)
\bibAnnoteFile{hutter2019automated}

\bibitem[{Jin et~al.(2019)Jin, Song, and Hu}]{jin2019auto}
Jin, H., Song, Q., and Hu, X. (2019).
\newblock Auto-keras: An efficient neural architecture search system.
\newblock In \emph{Proceedings of the 25th ACM SIGKDD international conference on knowledge discovery \& data mining}. 1946--1956
\bibAnnoteFile{jin2019auto}

\bibitem[{Kazemitabaar et~al.(2023)Kazemitabaar, Chow, Ma, Ericson, Weintrop, and Grossman}]{kazemitabaar2023studying}
Kazemitabaar, M., Chow, J., Ma, C. K.~T., Ericson, B.~J., Weintrop, D., and Grossman, T. (2023).
\newblock Studying the effect of ai code generators on supporting novice learners in introductory programming.
\newblock In \emph{Proceedings of the 2023 CHI Conference on Human Factors in Computing Systems}. 1--23
\bibAnnoteFile{kazemitabaar2023studying}

\bibitem[{Ke et~al.(2023{\natexlab{a}})Ke, Liu, Sun, Xue, Huang, Lu et~al.}]{ke2023artifact}
Ke, J., Liu, K., Sun, Y., Xue, Y., Huang, J., Lu, Y., et~al. (2023{\natexlab{a}}).
\newblock Artifact detection and restoration in histology images with stain-style and structural preservation.
\newblock \emph{IEEE Transactions on Medical Imaging} 42, 3487--3500
\bibAnnoteFile{ke2023artifact}

\bibitem[{Ke et~al.(2020{\natexlab{a}})Ke, Shen, Guo, Wright, Jing, and Liang}]{ke2020high}
Ke, J., Shen, Y., Guo, Y., Wright, J.~D., Jing, N., and Liang, X. (2020{\natexlab{a}}).
\newblock A high-throughput tumor location system with deep learning for colorectal cancer histopathology image.
\newblock In \emph{International Conference on Artificial Intelligence in Medicine} (Springer), 260--269
\bibAnnoteFile{ke2020high}

\bibitem[{Ke et~al.(2020{\natexlab{b}})Ke, Shen, Guo, Wright, and Liang}]{ke2020prediction}
Ke, J., Shen, Y., Guo, Y., Wright, J.~D., and Liang, X. (2020{\natexlab{b}}).
\newblock A prediction model of microsatellite status from histology images.
\newblock In \emph{Proceedings of the 2020 10th International Conference on Biomedical Engineering and Technology}. 334--338
\bibAnnoteFile{ke2020prediction}

\bibitem[{Ke et~al.(2023{\natexlab{b}})Ke, Shen, Lu, Guo, and Shen}]{ke2023mine}
Ke, J., Shen, Y., Lu, Y., Guo, Y., and Shen, D. (2023{\natexlab{b}}).
\newblock Mine local homogeneous representation by interaction information clustering with unsupervised learning in histopathology images.
\newblock \emph{Computer Methods and Programs in Biomedicine} 235, 107520
\bibAnnoteFile{ke2023mine}

\bibitem[{LeDell and Poirier(2020)}]{ledell2020h2o}
LeDell, E. and Poirier, S. (2020).
\newblock H2o automl: Scalable automatic machine learning.
\newblock In \emph{Proceedings of the AutoML Workshop at ICML} (ICML San Diego, CA, USA), vol. 2020
\bibAnnoteFile{ledell2020h2o}

\bibitem[{Liu et~al.(2024{\natexlab{a}})Liu, Zhou, Wang, Wang, Wang, Cao et~al.}]{liu2024toward}
Liu, X., Zhou, T., Wang, C., Wang, Y., Wang, Y., Cao, Q., et~al. (2024{\natexlab{a}}).
\newblock Toward the unification of generative and discriminative visual foundation model: A survey.
\newblock \emph{The Visual Computer} , 1--42
\bibAnnoteFile{liu2024toward}

\bibitem[{Liu et~al.(2024{\natexlab{b}})Liu, Chen, Wang, and Shen}]{liu2024autoproteinengine}
Liu, Y., Chen, Z., Wang, Y.~G., and Shen, Y. (2024{\natexlab{b}}).
\newblock Autoproteinengine: A large language model driven agent framework for multimodal automl in protein engineering.
\newblock \emph{arXiv preprint arXiv:2411.04440}
\bibAnnoteFile{liu2024autoproteinengine}

\bibitem[{Liu et~al.(2024{\natexlab{c}})Liu, Chen, Wang, and Shen}]{liu2024toursynbio}
Liu, Y., Chen, Z., Wang, Y.~G., and Shen, Y. (2024{\natexlab{c}}).
\newblock Toursynbio-search: A large language model driven agent framework for unified search method for protein engineering.
\newblock In \emph{2024 IEEE International Conference on Bioinformatics and Biomedicine (BIBM)} (IEEE), 5395--5400
\bibAnnoteFile{liu2024toursynbio}

\bibitem[{Luo et~al.(2024{\natexlab{a}})Luo, Feng, Nong, and Shen}]{luo2024autom3l}
Luo, D., Feng, C., Nong, Y., and Shen, Y. (2024{\natexlab{a}}).
\newblock Autom3l: An automated multimodal machine learning framework with large language models.
\newblock In \emph{Proceedings of the 32nd ACM International Conference on Multimedia}. 8586--8594
\bibAnnoteFile{luo2024autom3l}

\bibitem[{Luo et~al.(2024{\natexlab{b}})Luo, Feng, Shen, and Ma}]{luo2024learning}
Luo, S., Feng, J., Shen, Y., and Ma, Q. (2024{\natexlab{b}}).
\newblock Learning to predict the optimal template in stain normalization for histology image analysis.
\newblock In \emph{International Conference on Artificial Intelligence in Medicine} (Springer), 95--103
\bibAnnoteFile{luo2024learning}

\bibitem[{Mantovani et~al.(2016)Mantovani, Horv{\'a}th, Cerri, Vanschoren, and De~Carvalho}]{mantovani2016hyper}
Mantovani, R.~G., Horv{\'a}th, T., Cerri, R., Vanschoren, J., and De~Carvalho, A.~C. (2016).
\newblock Hyper-parameter tuning of a decision tree induction algorithm.
\newblock In \emph{2016 5th Brazilian Conference on Intelligent Systems (BRACIS)} (IEEE), 37--42
\bibAnnoteFile{mantovani2016hyper}

\bibitem[{Miah and Zhu(2024)}]{miah2024user}
Miah, T. and Zhu, H. (2024).
\newblock User centric evaluation of code generation tools.
\newblock In \emph{2024 IEEE International Conference on Artificial Intelligence Testing (AITest)} (IEEE), 109--119
\bibAnnoteFile{miah2024user}

\bibitem[{Olson and Moore(2016)}]{olson2016tpot}
Olson, R.~S. and Moore, J.~H. (2016).
\newblock Tpot: A tree-based pipeline optimization tool for automating machine learning.
\newblock In \emph{Workshop on automatic machine learning} (PMLR), 66--74
\bibAnnoteFile{olson2016tpot}

\bibitem[{Patibandla et~al.(2021)Patibandla, Srinivas, Mohanty, and Pattanaik}]{patibandla2021automatic}
Patibandla, R.~L., Srinivas, V.~S., Mohanty, S.~N., and Pattanaik, C.~R. (2021).
\newblock Automatic machine learning: An exploratory review.
\newblock In \emph{2021 9th International Conference on Reliability, Infocom Technologies and Optimization (Trends and Future Directions)(ICRITO)} (IEEE), 1--9
\bibAnnoteFile{patibandla2021automatic}

\bibitem[{Pham et~al.(2018)Pham, Guan, Zoph, Le, and Dean}]{pham2018efficient}
Pham, H., Guan, M., Zoph, B., Le, Q., and Dean, J. (2018).
\newblock Efficient neural architecture search via parameters sharing.
\newblock In \emph{International conference on machine learning} (PMLR), 4095--4104
\bibAnnoteFile{pham2018efficient}

\bibitem[{Sanders and Giraud-Carrier(2017)}]{sanders2017informing}
Sanders, S. and Giraud-Carrier, C. (2017).
\newblock Informing the use of hyperparameter optimization through metalearning.
\newblock In \emph{2017 IEEE International Conference on Data Mining (ICDM)} (IEEE), 1051--1056
\bibAnnoteFile{sanders2017informing}

\bibitem[{Shen(2024)}]{shen2024knowledgeie}
Shen, Y. (2024).
\newblock Knowledgeie: Unifying online-offline distillation based on knowledge inheritance and evolution.
\newblock In \emph{2024 International Joint Conference on Neural Networks (IJCNN)} (IEEE), 1--8
\bibAnnoteFile{shen2024knowledgeie}

\bibitem[{Shen et~al.(2024{\natexlab{a}})Shen, Chen, Mamalakis, Liu, Li, Su et~al.}]{shen2024toursynbio}
Shen, Y., Chen, Z., Mamalakis, M., Liu, Y., Li, T., Su, Y., et~al. (2024{\natexlab{a}}).
\newblock Toursynbio: A multi-modal large model and agent framework to bridge text and protein sequences for protein engineering.
\newblock In \emph{2024 IEEE International Conference on Bioinformatics and Biomedicine (BIBM)} (IEEE), 2382--2389
\bibAnnoteFile{shen2024toursynbio}

\bibitem[{Shen et~al.(2023)Shen, Guo, Wu, Huang, Le, Zhou et~al.}]{shen2023movit}
Shen, Y., Guo, P., Wu, J., Huang, Q., Le, N., Zhou, J., et~al. (2023).
\newblock Movit: Memorizing vision transformers for medical image analysis.
\newblock In \emph{International Workshop on Machine Learning in Medical Imaging} (Springer), 205--213
\bibAnnoteFile{shen2023movit}

\bibitem[{Shen et~al.(2024{\natexlab{b}})Shen, He, and Unberath}]{shen2024promptable}
Shen, Y., He, G., and Unberath, M. (2024{\natexlab{b}}).
\newblock Promptable counterfactual diffusion model for unified brain tumor segmentation and generation with mris.
\newblock In \emph{International Workshop on Foundation Models for General Medical AI} (Springer), 81--90
\bibAnnoteFile{shen2024promptable}

\bibitem[{Shen et~al.(2025{\natexlab{a}})Shen, Li, Liu, Li, Porras, and Unberath}]{shen2025operating}
Shen, Y., Li, C., Liu, B., Li, C.-Y., Porras, T., and Unberath, M. (2025{\natexlab{a}}).
\newblock Operating room workflow analysis via reasoning segmentation over digital twins.
\newblock \emph{arXiv preprint arXiv:2503.21054}
\bibAnnoteFile{shen2025operating}

\bibitem[{Shen et~al.(2024{\natexlab{c}})Shen, Li, Shao, Inigo~Romillo, Jindal, Dreizin et~al.}]{shen2024fastsam3d}
Shen, Y., Li, J., Shao, X., Inigo~Romillo, B., Jindal, A., Dreizin, D., et~al. (2024{\natexlab{c}}).
\newblock Fastsam3d: An efficient segment anything model for 3d volumetric medical images.
\newblock In \emph{International Conference on Medical Image Computing and Computer-Assisted Intervention} (Springer), 542--552
\bibAnnoteFile{shen2024fastsam3d}

\bibitem[{Shen et~al.(2025{\natexlab{b}})Shen, Liu, Li, Seenivasan, and Unberath}]{shen2025online}
Shen, Y., Liu, B., Li, C., Seenivasan, L., and Unberath, M. (2025{\natexlab{b}}).
\newblock Online reasoning video segmentation with just-in-time digital twins.
\newblock \emph{arXiv preprint arXiv:2503.21056}
\bibAnnoteFile{shen2025online}

\bibitem[{Shen et~al.(2024{\natexlab{d}})Shen, Lv, Zhu, and Wang}]{shen2024proteinengine}
Shen, Y., Lv, O., Zhu, H., and Wang, Y.~G. (2024{\natexlab{d}}).
\newblock Proteinengine: Empower llm with domain knowledge for protein engineering.
\newblock In \emph{International Conference on Artificial Intelligence in Medicine} (Springer), 373--383
\bibAnnoteFile{shen2024proteinengine}

\bibitem[{Shen et~al.(2022)Shen, Xu, Yang, Li, and Guo}]{shen2022self}
Shen, Y., Xu, L., Yang, Y., Li, Y., and Guo, Y. (2022).
\newblock Self-distillation from the last mini-batch for consistency regularization.
\newblock In \emph{Proceedings of the IEEE/CVF conference on computer vision and pattern recognition}. 11943--11952
\bibAnnoteFile{shen2022self}

\bibitem[{Socher et~al.(2013)Socher, Perelygin, Wu, Chuang, Manning, Ng et~al.}]{socher2013recursive}
Socher, R., Perelygin, A., Wu, J., Chuang, J., Manning, C.~D., Ng, A.~Y., et~al. (2013).
\newblock Recursive deep models for semantic compositionality over a sentiment treebank.
\newblock In \emph{Proceedings of the 2013 conference on empirical methods in natural language processing}. 1631--1642
\bibAnnoteFile{socher2013recursive}

\bibitem[{Sun et~al.(2023)Sun, Song, Gui, Ma, and Wang}]{sun2023automl}
Sun, Y., Song, Q., Gui, X., Ma, F., and Wang, T. (2023).
\newblock Automl in the wild: Obstacles, workarounds, and expectations.
\newblock In \emph{Proceedings of the 2023 CHI Conference on Human Factors in Computing Systems}. 1--15
\bibAnnoteFile{sun2023automl}

\bibitem[{Tambon et~al.(2025)Tambon, Moradi-Dakhel, Nikanjam, Khomh, Desmarais, and Antoniol}]{tambon2025bugs}
Tambon, F., Moradi-Dakhel, A., Nikanjam, A., Khomh, F., Desmarais, M.~C., and Antoniol, G. (2025).
\newblock Bugs in large language models generated code: An empirical study.
\newblock \emph{Empirical Software Engineering} 30, 1--48
\bibAnnoteFile{tambon2025bugs}

\bibitem[{Trirat et~al.(2024)Trirat, Jeong, and Hwang}]{trirat2024automl}
Trirat, P., Jeong, W., and Hwang, S.~J. (2024).
\newblock Automl-agent: A multi-agent llm framework for full-pipeline automl.
\newblock \emph{arXiv preprint arXiv:2410.02958}
\bibAnnoteFile{trirat2024automl}

\bibitem[{Tsai et~al.(2023)Tsai, Tsai, Huang, Yang, and Lin}]{tsai2023automl}
Tsai, Y.-D., Tsai, Y.-C., Huang, B.-W., Yang, C.-P., and Lin, S.-D. (2023).
\newblock Automl-gpt: Large language model for automl.
\newblock \emph{arXiv preprint arXiv:2309.01125}
\bibAnnoteFile{tsai2023automl}

\bibitem[{Wang et~al.(2024)Wang, He, He, Ye, and Shen}]{wang2024histology}
Wang, C., He, Z., He, J., Ye, J., and Shen, Y. (2024).
\newblock Histology image artifact restoration with lightweight transformer based diffusion model.
\newblock In \emph{International Conference on Artificial Intelligence in Medicine} (Springer), 81--89
\bibAnnoteFile{wang2024histology}

\bibitem[{Wang and Shen(2024)}]{wang2024evaluating}
Wang, L. and Shen, Y. (2024).
\newblock Evaluating causal reasoning capabilities of large language models: A systematic analysis across three scenarios.
\newblock \emph{Electronics} 13, 4584
\bibAnnoteFile{wang2024evaluating}

\bibitem[{Wen et~al.(2024)Wen, Wang, Yi, Ke, and Shen}]{wen2024diffimpute}
Wen, Y., Wang, Y., Yi, K., Ke, J., and Shen, Y. (2024).
\newblock Diffimpute: Tabular data imputation with denoising diffusion probabilistic model.
\newblock In \emph{2024 IEEE International Conference on Multimedia and Expo (ICME)} (IEEE), 1--6
\bibAnnoteFile{wen2024diffimpute}

\bibitem[{Zan et~al.(2022)Zan, Chen, Zhang, Lu, Wu, Guan et~al.}]{zan2022large}
Zan, D., Chen, B., Zhang, F., Lu, D., Wu, B., Guan, B., et~al. (2022).
\newblock Large language models meet nl2code: A survey.
\newblock \emph{arXiv preprint arXiv:2212.09420}
\bibAnnoteFile{zan2022large}

\bibitem[{Zhang et~al.(2023)Zhang, Gong, Wu, Liu, and Zhou}]{zhang2023automl}
Zhang, S., Gong, C., Wu, L., Liu, X., and Zhou, M. (2023).
\newblock Automl-gpt: Automatic machine learning with gpt.
\newblock \emph{arXiv preprint arXiv:2305.02499}
\bibAnnoteFile{zhang2023automl}

\bibitem[{Zimmer et~al.(2021)Zimmer, Lindauer, and Hutter}]{zimmer2021auto}
Zimmer, L., Lindauer, M., and Hutter, F. (2021).
\newblock Auto-pytorch: Multi-fidelity metalearning for efficient and robust autodl.
\newblock \emph{IEEE transactions on pattern analysis and machine intelligence} 43, 3079--3090
\bibAnnoteFile{zimmer2021auto}

\bibitem[{Zoph(2016)}]{zoph2016neural}
Zoph, B. (2016).
\newblock Neural architecture search with reinforcement learning.
\newblock \emph{arXiv preprint arXiv:1611.01578}
\bibAnnoteFile{zoph2016neural}

\end{thebibliography}


\end{document}